\documentclass[11pt]{article}
\usepackage{algorithm}
\usepackage{authblk}
\usepackage{appendix}
\usepackage{amsmath}
\usepackage[sort&compress,authoryear,round]{natbib}
\usepackage{url}
\usepackage{array}
\usepackage{algpseudocode}
\usepackage[dvipsnames]{xcolor}
\usepackage[colorlinks=true,				
pdfborder=1 1 1,				
linkcolor=Sepia,					
citecolor=Sepia,					
urlcolor=Sepia,					
]{hyperref}

\newcommand{\Eco}{\textit{E.~coli}}
\newcommand{\Ecol}{\textit{Escherichia coli}}
\newcommand{\Mtu}{\textit{M.~tuberculosis}}
\newcommand{\Sty}{\textit{S.~}typhimurium}
\newcommand{\fastSL}{\textsc{Fast-SL}}

\begin{document}
\title{\fastSL: An efficient algorithm to identify synthetic lethal reaction sets in metabolic networks}
\author[1]{Aditya Pratapa}
\affil[1]{\it Department of Biotechnology, Bhupat and Jyoti Mehta School of Biosciences\\
\sc Indian Institute of Technology Madras\\
Chennai -- 600 036, INDIA}
\author[2]{Shankar Balachandran}
\affil[2]{\it Department of Computer Science and Engineering, \sc Indian Institute of Technology Madras}
\author[1]{Karthik Raman\thanks{Corresponding author. E-mail: kraman@iitm.ac.in}}
\date{}
\maketitle

\begin{abstract}
Synthetic lethal reaction/gene sets are sets of reactions/genes where only the simultaneous removal of all reactions/genes in the set abolishes growth of an organism.  \textit{In silico}, synthetic lethal sets can be identified by simulating the effect of removal of gene sets from the reconstructed genome-scale metabolic network of an organism. Flux balance analysis (FBA), based on linear programming, has emerged as a powerful tool for the  \textit{in silico} analyses of metabolic networks. To identify all possible synthetic lethal reactions combinations, an exhaustive sampling of all possible combinations is computationally expensive. We surmount the computational complexity of exhaustive search by iteratively restricting the sample space of reaction combinations for search, resulting in a substantial reduction in the running time. We here propose an algorithm, \fastSL, which provides an efficient way to analyse metabolic networks for higher order lethal reaction sets. \fastSL\ offers a substantial speed-up through a massive reduction in the search space for synthetic lethals; in the case of \Eco, \fastSL\ reduces the search space for synthetic lethal triplets by over 4000-fold. \fastSL\ also compares favourably with SL~Finder, an algorithm for identifying synthetic lethal sets, by ~\cite{Suthers2009SLFinder}, which involves the solution of a bi-level Mixed Integer Linear Programming problem. We have implemented the algorithm in MATLAB, building upon COBRA toolbox v2.0.

\end{abstract}

\clearpage
\section{Introduction}
In the last decade, genome-scale metabolic networks have been reconstructed for many organisms. These networks have been studied using tools such as Flux Balance Analysis (FBA)~\citep{Varma1994Stoichiometric,Kauffman2003Advances}, for the identification of drug targets~\citep{Jamshidi2007Investigating,Thiele2011Community}, or targets for metabolic engineering~\citep{Alper2005Identifying}. More recent studies have identified combinations of genes, which when simultaneously deleted, abolish growth \textit{in silico}~\citep{Suthers2009SLFinder}. These sets, termed `synthetic lethals', reveal complex interactions in metabolic networks. Synthetic lethals have been analysed in the past for prediction of novel genetic interactions and analysing the extent of robustness of biological networks~\citep{Raghunathan2009Constraintbased}. 

Enumeration of synthetic lethals of higher orders was previously performed for the metabolic network of yeast through an exhaustive search, by parallelising the deletions on a cluster of computers~\citep{Deutscher2006Multiple}. However, exhaustive enumeration is computationally very expensive, and prohibitive, in case of large metabolic networks. Another algorithm for identifying synthetic lethals is `SL~Finder', published by Maranas and co-workers~\citeyearpar{Suthers2009SLFinder}. SL~Finder elegantly poses the identification of synthetic lethals as a bi-level Mixed Integer Linear Programming (MILP) problem; the algorithm has been applied for the identification of synthetic lethal doublets and triplets in \Ecol. We here propose an alternative algorithm, \fastSL, which circumvents the computational complexity of both exhaustive enumeration and bi-level MILP, through an iterative reduction of the search space for single, double and triple lethal reaction sets. \fastSL\  finds application in the identification of combinatorial drug targets and novel genetic interactions.

\section{Algorithm}

\subsection{Overview}
FBA has been previously extended to identify synthetic lethals, either by exhaustive search~\citep{Deutscher2006Multiple} or by targeted enumeration (SL Finder; \citet{Suthers2009SLFinder}). FBA involves the formulation of a Linear Programming (LP) problem, whose objective function typically is to maximise flux through the biomass reaction ($v_{bio}$), subject to the constraints obtained from the stoichiometry of the metabolic network (represented by the stoichiometric matrix $\textbf{S}$). FBA has also be used to simulate the effects of the removal of one or more genes/reactions from a metabolic network. The phenotype obtained as a result of gene/reaction deletion is classified as a lethal phenotype, if the maximum growth-rate obtained by FBA is less than 1\% of the \textit{in silico} maximum wild-type growth rate~\citep{Deutscher2008Can}. We here propose an alternative algorithm to identify synthetic lethals, using an iterative approach that greatly reduces the search space for the synthetic lethals. 

\subsection{\fastSL\ Algorithm}

The objective of \fastSL\ is to enumerate combinations of reactions, which when deleted, abolish growth. We achieve this by a combination of pruning the search space and exhaustively iterating through the remaining combinations.  We successively compute $J_{sl}$, the set of single lethal reactions, $J_{dl}$, the set of synthetic lethal reaction pairs, and $J_{tl}$, the set of synthetic lethal reaction triplets. Initially, we use FBA to compute a flux distribution, corresponding to maximum growth rate, while minimising the sum of absolute values of the fluxes (the $L_1$-norm of the flux vector). We hereafter denote this flux distribution as the `minNorm' solution of the FBA LP problem. We denote the set of reactions that carry a flux in this minNorm solution as $J_{nz}$. $J_{nz}$ is a minimal representative set of reactions, necessary for the organism to sustain growth.

Reactions are essential for the growth of an organism if they cannot be bypassed, i.e. they have no alternate in the system. Some reactions have one or more alternates; it is also possible that one set of reactions (a pathway) is replaceable by another set, perhaps not even of the same size. We argue that $J_{nz}$  contains all single lethal reactions, as well as at least one reaction from each set of reactions that is essential for growth (see Appendix~A for further details). In other words, this set of reactions contains $J_{sl}$ and at least one reaction from each lethal reaction pair, triplet etc. We compute $J_{sl}$ by performing exhaustive single reaction deletions in $J_{nz}$. The algorithm for identification of lethal phenotypes of order two involves two phases:
\begin{description}
\item[Phase 1] For every $v_i \in J_{nz} - J_{sl}$, removal of $v_i$ redirects the corresponding minNorm flux  $J_{nz,i}$, through a new set of reactions denoted by \mbox{$J_{nz,i} - J_{nz}$}, which is the new search space for the lethal counterpart of $v_i$. 
\item[Phase 2] We exhaustively search for lethal pairs in the set $J_{nz}-J_{sl}$. Algorithm~1 shows the algorithm for enumerating synthetic lethal sets of order up to two. The algorithm for enumerating synthetic lethal triple reaction sets extends this idea further, as detailed in Appendix~B.
\end{description}

\begin{algorithm}[H]
\small
\begin{algorithmic}
\State \textbf{Input:} SBML model of an organism
\State \textbf{Output:} Set of single lethal reactions $J_{sl}$, Set of double lethal reactions $J_{dl}$\\
\State \textbf{Phase 1} 
\State  Perform FBA to obtain the minNorm maximum wild-type growth rate, $v_{bio}$
\State  Identify set of reactions $J_{nz}$, having non-zero fluxes 
\State  Set $0.01 \ast v_{bio}$ as the cut-off for lethality, $v_{co}$
\For {each reaction $i \in J_{nz}$}
\State Set the upper and lower bounds of $v_i$ to zero
\State Perform FBA to obtain minNorm solution $v_{bio,i}$

\If {$v_{bio,i} \leq v_{co}$}
	\State Add $i$ to the set $J_{sl}$
\Else 
\State Identify set of reactions $J_{nz,i}$, having non-zero fluxes
	\For {each  $j \in J_{nz,i}-J_{nz}$}
	\State Set the upper and lower bounds of $v_j$ to zero
	\State Perform FBA to maximise growth rate $v_{bio,ij}$
	\If {$v_{bio,ij} \leq v_{co}$}
\State Add $\left\{i,j \right\}$ to the set $J_{dl}$

\EndIf
\State Reset the bounds on $v_j$
\EndFor
\EndIf
\State Reset bounds on $v_i$
\EndFor
\\
\State \textbf{Phase 2} 

\For {each reaction pair $\lbrace i,j \rbrace \in J_{nz}-J_{sl}$ such that $i\neq j$}
\State Set the upper and lower bounds of $v_i$ and $v_j$ to zero
\State Perform FBA to maximise growth rate $v_{bio,ij}$
\If {$v_{bio,ij} \leq v_{co}$}
\State Add $\lbrace i,j\rbrace$ to the set $J_{dl}$
\EndIf
\State Reset the bounds on $v_i$ and $v_j$
\EndFor

\end{algorithmic}
\caption{\textbf{Algorithm to identify single and double lethal reaction sets.}}
\end{algorithm}

\section{Results and Discussion}
We performed reaction deletions up to the order of three on  genome-scale metabolic networks of \textit{Escherichia coli} \textit{iAF}1260~\citep{Feist2007Genomescale}, \textit{Salmo\-nella enterica} Typhimurium \textit{iIT}1176~\citep{Thiele2011Community} and \textit{Mycobacterium tuberculosis iNJ}661~\citep{Jamshidi2007Investigating} using COBRA Toolbox v2.0~\citep{Schellenberger2011Quantitative} on MATLAB R2013b (Mathworks Inc.) and the Gurobi solver (v5.6.3, Gurobi Inc.).

Table~\ref{tab:LPstats} enumerates the number of LPs solved  by the proposed algorithm as compared to the exhaustive enumeration. For \Eco\ and \Sty, we observe more than a 2000-fold reduction in the search space for synthetic lethal sets; for the smaller model of \Mtu, it is over 200-fold. We have identified 96 synthetic lethal reaction pairs and 247 lethal reaction triplets in the \textit{E.~coli iAF}1260 model. A complete listing of lethal reaction sets for all three organisms is available in \emph{Supplementary File~1}. Synthetic lethality can be further extended to quadruples and other higher orders using a similar approach.  The concept of synthetic lethal reaction sets can be easily extended to lethal gene sets using the gene--protein--reaction associations, which provide details on which genes encode for which proteins, and the reactions catalysed by these proteins.

\begin{table*}[t]
\centering
\begin{tabular}{l>{\raggedleft\arraybackslash}p{2.4cm}>{\raggedleft\arraybackslash}p{2.4cm}>{\raggedleft\arraybackslash}p{2.4cm}}
\hline
\textbf{Model} & \textbf{Number of Reactions} & \textbf{Exhaustive LPs} & \textbf{LPs solved by \fastSL}\\\hline
\textit{iAF}1260 & {\raggedright $2,382$} & $9.27 \times 10^8$ & $229,938$\\
\textit{iIT}1176 & $2,546$ & $1.04 \times 10^9$ & $470,404$\\
\textit{iNJ}661 & $1,028$  & $6.17 \times 10^7$ & $182,457$\\
\hline
\end{tabular}

\caption{\textbf{Number of LP Problems to be evaluated as compared to exhaustive enumeration of synthetic lethal reactions.} For each of the three models, \fastSL\ presents a significant reduction in search space.}
\label{tab:LPstats}
\end{table*}

 Table~\ref{tab:exhaustive_vs_fastSL} illustrates the vast improvement in computational time over an exhaustive search. \citet{Suthers2009SLFinder} report that their algorithm is able to enumerate all synthetic lethal triple reaction sets in $\approx 6.75$ days, on a 3~GHz processor. We have been unable to perform a systematic comparison owing to the difference in platforms (General Algebraic Modeling System (GAMS) vs MATLAB), as well as processors used. However, we note that the savings obtained through a pruning of reaction space and the fact that we solve only a large number of small LPs instead of a bi-level MILP, render \fastSL\ as a powerful alternative, for metabolic networks of any size.
 
\begin{table*}[h]
\centering
\begin{tabular}{p{1.5cm}>{\raggedleft\arraybackslash}p{2cm}>{\raggedleft\arraybackslash}p{2.4cm}>{\raggedleft\arraybackslash}p{2cm}r@{\hspace{1ex}}l}
\hline
\textbf{Order of SLs} & \textbf{Exhaustive LPs} & \textbf{CPU time (estimated)} & \textbf{LPs solved by \fastSL} & \multicolumn{2}{r}{\textbf{CPU time}} \\
\hline
Single & $2.05 \times 10^3$ & $\approx 150.8$ s & $379$ & $\approx 25.1$& s\\
Double& $1.57 \times 10^6$ & $\approx 34.3$ h & $6,084$ & $\approx 8.4$ &min\\
Triple & $9.27 \times 10^8$ & $\approx 817.5$ d & $223,469$ & $\approx 6.27$ &h\\
\hline
\end{tabular}
\caption{\textbf{Comparison of the number for LPs to be solved and CPU time for \textit{E.~coli}  for exhaustive enumeration versus the \fastSL\ algorithm}. The times reported are for a workstation with a 2.4GHz Intel Xeon E5645 processor and 16GB of DDR3 RAM.}
\label{tab:exhaustive_vs_fastSL}
\end{table*}

Overall, \fastSL\ presents a massive reduction in the search space over an exhaustive enumeration approach and the `SL Finder' algorithm. Our approach also is inherently parallelisable, which can lead to further savings in computational time.

\section*{Supplementary Material}

\paragraph{Supplementary File 1} This XLS file contains a complete listing of (a)~single lethal reactions, (b)~synthetic lethal reaction pairs and (c)~synthetic lethal reaction triplets for \textit{Escherichia coli} \textit{iAF}1260, \textit{Salmonella enterica} Typhimurium \textit{iIT}1176  and \textit{Mycobacterium tuberculosis iNJ}661.

\mbox{}\\
\textbf{Availability:} The MATLAB implementation of the algorithm is available at \url{https://home.iitm.ac.in/kraman/lab/research/fast-sl}

\subsection*{Acknowledgements}
The authors would like to thank Aarthi Ravikrishnan for useful discussions. \\
\textit{Funding}: KR acknowledges funding from IIT Madras and the grant BT/ PR4949/BRB/10/1048/2012 from the Department of Biotechnology, Government of India.\\
\\
\textit{Conflict of Interest}: none declared.


\pagebreak
\appendixtitleon
\appendixtitletocon
\begin{appendices}
\section {Fast SL Algorithm}
To identify synthetic lethal reactions from the given reaction set $J$, we compute the minNorm solution using Flux Balance Analysis, as follows: 

\begin{align}
&\text{min.   } 
\Sigma_j \vert v_j \vert \\
\text{subject to:} \notag \\
& v_{biomass} = v_{biomass,max} \\
& \Sigma_j s_{ij}v_j= 0 & & \forall i \in M \\
& LB_j \leq v_j \leq UB_j & & \forall j \in J
\end{align}
where,\\
$J$ represents the set of all reactions in the metabolic network\\
$M$ represents the set of all metabolites in the metabolic network\\
$v_j$ represents the flux through the $j^{th}$ reaction \\
$v_{biomass}$ represents the flux through the biomass  reaction\\
$v_{biomass,max}$ represents the maximum biomass flux obtained using FBA\\
$s_{ij}$ represent the $ij^{th}$ element in stoichiometric matrix $\textbf{S}$\\
$LB_j$  and $UB_j$ represent the lower and upper bounds of the fluxes through the $j^{th}$ reaction\\

This can be computed by setting the minNorm flag to `one' using the COBRA Toolbox in MATLAB. Reactions that carry a non-zero flux are represented by $J_{nz}$. The set $J_{nz}$ is a minimal representative set of reactions necessary for the organism to sustain growth. For example, the \textit{iAF}1260 model of \Eco\ has $2,382$ reactions. The formulation above produces 406 reactions in the set $J_{nz}$ for aerobic growth on minimal glucose conditions (the wild-type growth rate is \mbox{$0.9290$ mmol gDW$^{-1}$ h$^{-1}$}).\\

\subsection*{Identifying Synthetic Lethal Reactions  ($J_{sl}$)}
\textbf{The set $J_{sl}$ is entirely contained in $J_{nz}$}. This is because, any reaction from $J_{sl}$, when constrained to zero flux, cannot sustain growth (definition of a lethal reaction). Conversely, any reaction from $J_{sl}$ cannot have zero flux when constrained to wild-type growth rate (given by constraint (2)) and hence belongs to $J_{nz}$. So, instead of analysing $J$ exhaustively ($\vert J \vert= 2081$ reactions for \textit{E.~coli}), it would suffice to analyse the 406 reactions of $J_{nz}$. This gives us 278 single lethal reactions, $J_{sl}$ (Note: Exchange reactions are not considered).

\subsection*{Identifying Synthetic Lethal Reaction Pairs  ($J_{dl}$)}
\textbf{For every pair in the set $J_{dl}$, at least one of the reaction will be in $J_{nz}$}. To understand this, consider any pair of reactions $i$ and $j \in J$. Only three types of such pairs exist:
\begin{itemize}\itemsep1pt
\item[\textbf{(i)}] $i$ and $j \not \in J_{nz}$
\item[\textbf{(ii)}] One of $i$ or $j \in J_{nz}$
\item[\textbf{(iii)}] Both $i$ and $j \in J_{nz}$
\end{itemize}
\textit{For pairs of type (i)}:\\
Suppose, both reactions of a pair do not belong to $J_{nz}$. This implies that both have zero flux through them under the minNorm formulation and hence constraining them simultaneously does not produce a lethal phenotype and therefore, is not a lethal pair. Our algorithm eliminates such pairs for consideration and hence reduces the search space substantially. In case of \textit{E.coli iAF}1260 model,the exhaustive search space has $^{2051-278}C_{2}=$ 1.57 million combinations for double reaction deletions analysis. Out of these, the pairs of type (i) correspond to 1.39 million and hence are eliminated.\\
\\
\textit{For pairs of type (ii) and (iii)}:\\
If at least one of the reaction pair belongs to $J_{nz}$, this can be a lethal pair. To analyse such pairs, we proposed two phases of analyses:\\
\textbf{ Phase 1}: We remove one reaction $i \in J_{nz}-J_{sl}$ at a time and solve for minNorm flux distribution and consider the new set of non-zero fluxes  $J_{nz,i}$. As this reaction deletion is not (single) lethal, to sustain growth, it redirects the flux through a new set of reactions that were previously not in $J_{nz}$, and can potentially be the lethal counterpart of  reaction $i$. Therefore, it would suffice to look into the reactions $j \in J_{nz,i}-J_{nz}$ exhaustively for simultaneous deletions that can potentially produce a lethal pair $\lbrace i,j\rbrace$.  This corresponds to pair of type (ii). Reducing the search space in this manner also leads to the reduction of search space (here 852 simulations are performed instead of 168,872 simulations).\\
\textbf{Phase 2}: We exhaustively look for lethal combinations for reactions \linebreak[4]$\lbrace i,j \rbrace \in J_{nz}-J_{sl}$. This corresponds to pairs of type (iii).

\pagebreak[4]
\section{Algorithm to find Lethal Reaction Triplets}
\fastSL\ can be further extended to identify synthetic lethal reaction triplets, $J_{tl}$, using a reduction procedure as given below:\\
\begin{algorithm}[ht]
\small
\begin{algorithmic}
\State \textbf{Input:} SBML model of an organism, $J_{sl}$, $J_{dl}$
\State \textbf{Output:} Lethal Reaction Triplets -- $J_{tl}$ \\

\State \textbf{Phase 1} 
\For {each reaction $i \in J_{nz}-J_{sl}$}
\State Set the upper and lower bounds of $v_i$ to zero
\State Perform FBA to obtain minNorm solution $v_{bio,i}$
\State Identify set of reactions $J_{nz,i}$, having non-zero fluxes
\For {each $j \in J_{nz,i}-J_{nz}$}
\If {$\lbrace i,j\rbrace \not \in J_{dl}$}
	\State Set the upper and lower bounds of $v_j$ to zero
	\State Perform FBA to obtain minNorm solution $v_{bio,ij}$
	\State Identify set of reactions $J_{nz,ij}$, having non-zero fluxes
	\For {each $k \in J_{nz,ij}-J_{nz}$}
\If {$\lbrace i,k \rbrace \not \in J_{dl}$ and $\lbrace j,k \rbrace \not \in J_{dl}$}
\State Set the upper and lower bounds of $v_j$ to zero
	\State Perform FBA to maximise growth rate $v_{bio,ijk}$
\If {$v_{bio,ijk} \leq v_{co}$}
\State Add $\lbrace i,j,k\rbrace$ to the set $J_{tl}$
\EndIf
\State Reset the bounds on $v_k$
\EndIf
\EndFor
\State Reset bounds on $v_j$
\EndIf
\EndFor
\State Reset bounds on $v_i$
\EndFor\\

	\algstore{myalg}

\caption{Algorithm to identify triple lethal reaction sets}
\end{algorithmic}
\end{algorithm}

\begin{algorithm}[ht]
\begin{algorithmic}
\algrestore{myalg}
\State \textbf{Phase 2} 
\For {each reaction pair $\lbrace i,j \rbrace \in J_{nz}-J_{sl}$ and $\not \in J_{dl}$}
\State Set the upper and lower bounds of $v_i$ and $v_j$ to zero
	\State Perform FBA to obtain minNorm solution $v_{bio,ij}$
	\State Identify set of reactions $J_{nz,ij}$, having non-zero fluxes
	
\For {each $v_k \in J_{nz,ij}-J_{nz}$}
\If {$\lbrace i,k \rbrace \not \in J_{dl}$ and $\lbrace j,k \rbrace \not \in J_{dl}$}
\State Set the upper and lower bounds of $v_k$ to zero
	\State Perform FBA to maximise growth rate $v_{bio,ijk}$
\If {$v_{bio,ijk} \leq v_{co}$}
\State Add $\lbrace i,j,k\rbrace$ to the set $J_{tl}$
\EndIf
\State Reset the bounds on $v_k$
\EndIf
\EndFor
\State Reset bounds on $v_i$ and $v_j$ 
\EndFor\\
\State \textbf{Phase 3} 
\For {each reaction triplet $\lbrace i,j,k \rbrace \in J_{nz}-J_{sl}$ such that $i\neq j\neq k$ and ($\lbrace i,j \rbrace$, $\lbrace j,k \rbrace$, $\lbrace i,k \rbrace$) $ \not \in J_{dl}$}
\State Set the upper and lower bounds of $v_i$, $v_j$ and $v_k$ to zero
\State Perform FBA to maximise growth rate $v_{bio,ijk}$
\If {$v_{bio,ijk} \leq v_{co}$}
\State Add $\lbrace i,j,k\rbrace$ to the set $J_{tl}$
\EndIf
\State Reset the bounds on $v_i$, $v_j$ and $v_k$
\EndFor
\end{algorithmic}
\end{algorithm}

\end{appendices}
\end{document}